# Simultaneous Mode and Wavelength Division Multiplexing On-Chip

Lian-Wee Luo[1,*], Noam Ophir[2,*], Christine Chen[2], Lucas H. Gabrielli[1], Carl B. Poitras[1], Keren Bergman[2] & Michal Lipson[1,3,#]

[1]*School of Electrical and Computer Engineering, Cornell University, Ithaca, NY 14853, USA*

[2]*Department of Electrical Engineering, Columbia University, New York, New York, USA*

[3]*Kavli Institute at Cornell for Nanoscale Science, Cornell University, Ithaca, NY 14853, USA*

[#]*ML292@cornell.edu*

*These authors contributed equally to this work.

**Abstract:** Significant effort in optical-fiber research has been put in recent years into realizing mode-division multiplexing (MDM) in conjunction with wavelength-division multiplexing (WDM) to enable further scaling of the communication bandwidth per fiber. In contrast almost all integrated photonics operate exclusively in the single-mode regime. MDM is rarely considered for integrated photonics due to the difficulty in coupling selectively to high-order modes which usually results in high inter-modal crosstalk. Here we show the first demonstration of simultaneous on-chip mode and wavelength division multiplexing with low modal crosstalk and loss. Our approach can potentially increase the aggregate data rate by many times for on-chip ultra-high bandwidth communications.

Current integrated photonics operate almost exclusively in the single-mode regime and utilize wavelength-division multiplexing (WDM)[1,2] which supports a limited scalability in bandwidth density. In contrast fiber communications is increasingly targeting multimode operation in conjunction with WDM to further scale the communication bandwidth transmitted per fiber[3]. Multimode communications in fibers have been demonstrated with space-division multiplexing (SDM) in multi-core fibers[4-8] or mode-division multiplexing (MDM) in few-mode fibers (FMF)[9-17] and have exploited each spatial mode as an independent channel. Here we show a platform enabling MDM in conjunction with WDM in integrated photonics for on-chip and chip-to-chip ultra-high bandwidth applications. This platform could increase the bandwidth density of on-chip interconnects, reduce the number of waveguide crossings for an on-chip interconnect, and add an additional design degree of freedom in future photonic networks.

Some of the key challenges of realizing on-chip MDM-enabled interconnects lie in creating mode (de)multiplexers with low modal crosstalk and loss which also support WDM (a key feature of many integrated-optics interconnect designs). Previous implementations of on-chip mode multiplexing based on Mach-Zehnder interferometers[18,19], Multi-Mode Interference (MMI) couplers[20-22], asymmetric directional couplers[23-25], and y-junctions[26-28] typically had large footprints, complex and strict design limitations, or only supported a limited number of optical modes. A compact and reconfigurable mode (de)multiplexer which can be straight-forwardly scaled to support numerous modes is essential for realizing MDM-WDM in integrated photonics.

We demonstrate on-chip MDM-WDM by engineering the propagation constants of high-confinement photonic structures in order to enable selective coupling to different spatial modes at different wavelengths. The silicon photonic platform is attractive for implementing this approach as the propagation constants of the different spatial modes can be engineered to differ significantly thanks to the high core-cladding (Si/SiO$_2$) index contrast. We choose a waveguide height for which the confinement is high and therefore widely different propagation constants can be achieved by varying the waveguide width. Figure 1a shows that for a given 250-nm tall silicon waveguide a large range of effective indices from 2.0 to 2.9 can be achieved corresponding to the propagation constants of the $TE_0$ through $TE_4$ spatial modes at λ = 1550 nm. Based on propagation constant matching, an optical mode in a single-mode waveguide can be evanescently coupled to a single spatial mode in an adjacent multimode waveguide, where the coupling strength to the mode will depend on the width of the multimode waveguide.

In order to realize a compact MDM device, we employ single-mode microring resonators to selectively couple to the different spatial modes in a multimode waveguide. In order to realize WDM capabilities, we design the free-spectral range (FSR) of the microrings to match the wavelength channel spacing. The device, which is designed to operate in TE mode, comprises of three identical microrings coupled to a multimode waveguide (see Fig. 1b). Each microring, made up of a 450-nm wide waveguide, is designed to support only the fundamental TE mode with an effective index of 2.46. The multimode bus waveguide comprises of several sections with tapering widths ranging from 450nm to 1.41 µm. When the bus waveguide width corresponds to 450 nm, 930 nm, or 1.41 µm, the effective indices of $TE_0$, $TE_1$, or $TE_2$ modes respectively match the effective index of the $TE_0$ mode of the microrings ($n_{eff}$ = 2.46) and therefore couple efficiently to the resonators. The three insets in Fig. 1b show such coupling of the $TE_0$ mode of the microring to the $TE_0$, $TE_1$, or $TE_2$ modes in the bus waveguide. Since the propagation loss in silicon ring is low, a low coupling strength (achievable with a short coupling length) at the two ring-waveguide coupling regions is sufficient to transfer all the power from the single-mode input waveguide to the multimode bus waveguide[29]. We design the ring resonance linewidth to be at least 15-GHz in order to enable 10-Gb/s data transmission with negligible signal degradation. We also include an integrated heater on top of each microring to tune the ring resonances to align to the WDM channels and thereby optimize the performance of the device[30]. This design can be easily modified to handle additional phase-matched modes by widening the multimode waveguide (Fig. 1a).

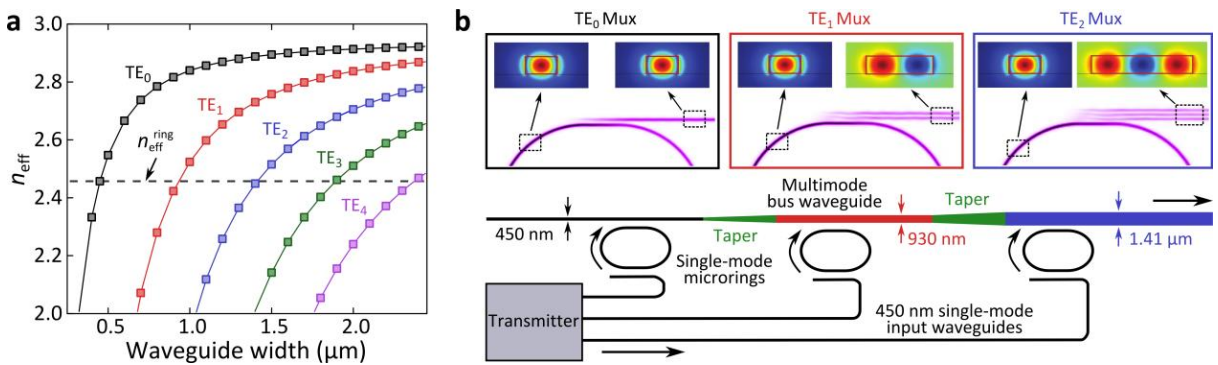

**Figure 1 | Phase matching condition for 250 nm tall silicon waveguides. a.** Simulated effective index of the optical modes in waveguide of different widths at λ = 1550 nm. **b.** Selective coupling of the single-mode microrings to a specific spatial mode in the multimode bus waveguide with each section of the multimode waveguide linked by adiabatic tapered waveguides. The insets show the selective coupling of each multiplexer ($TE_0$, $TE_1$ and $TE_2$).

We show that the fabricated mode multiplexers introduce crosstalk as low as -30 dB between the modes. The fabrication details for this MDM device are documented in the Methods section (see Fig. 2a). We image the optical modes at the output of the multimode waveguide to verify the excitation of the different spatial modes (see Methods for the imaging experimental details). We observe well-defined $TE_0$, $TE_1$, and $TE_2$ modes as predicted by simulation (Fig. 2b). From the spectral transmission scans for each combination of input and output ports, we can quantify the amount of crosstalk resulting from the spatial mode multiplexing and demultiplexing. Figure 2c shows the transmission spectrum at output port 1 (see port definitions in Fig. 2a) from each input. The insertion loss of this port is 13 dB and the optical crosstalk (defined as the ratio of desired signal power to the sum of the interfering channels' power) is as low as -30 dB. The insertion loss of port 2 is 16 dB and the optical crosstalk is -18 dB (Fig. 2d). The insertion loss of port 3 is 26 dB and crosstalk is -13 dB (Fig. 2e). The main contribution to the insertion loss is the aggregate 10-dB fiber-to-chip coupling loss. The rest of the insertion loss is attributed to the waveguide propagation loss and ring intrinsic loss. By ensuring critical coupling between the waveguides and rings, achievable on-chip losses of this device are expected to total around 1.5 dB. The higher insertion loss in port 3 compared to the other 2 ports is due to a suboptimal ring coupling gap. The crosstalk from the unwanted input signals can be minimized by optimizing the coupling length between the ring and the multimode waveguide to reduce coupling of undesired modes (Supplementary Fig. 1). We expect the crosstalk at output port 3 of the current device to be less than -16 dB for an optimized coupling length of 6 μm (larger than the fabricated one with only 5 μm) (see Fig. 2f). The crosstalk can be further reduced by introducing weaker coupling at the ring-multimode waveguide coupling region (by having a larger coupling gap) to lower the maximum coupling of the undesired modes at the expense of longer coupling length to maintain the critical coupling condition (Supplementary Fig. 2).

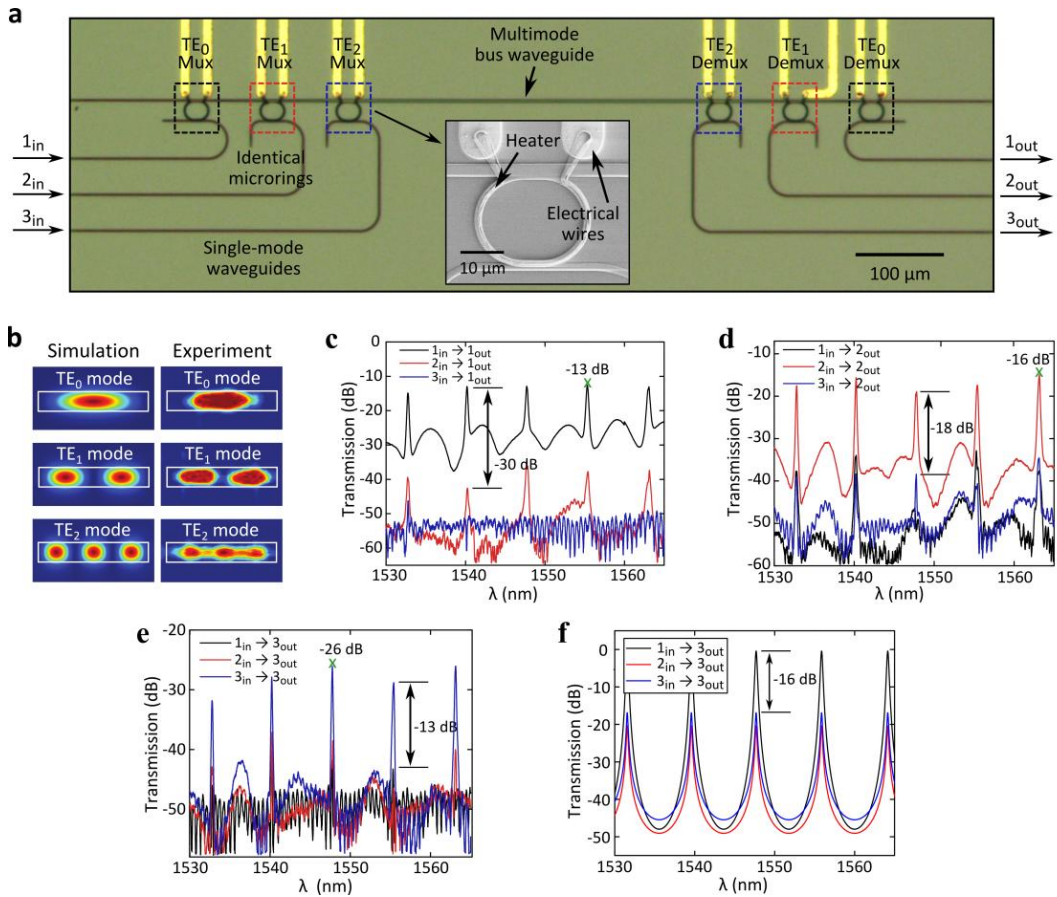

**Figure 2 | Optical performance of the fabricated device. a.** Microscope image of the fabricated device. Inset: SEM showing the heater to tune each individual ring resonator. **b.** Simulated and experimental images of the optical modes at the cross-section of the multimode waveguide. **c-e.** Optical transmission and crosstalk at the three output ports for signal injection on each of the three input ports. **f.** Simulated transmission and crosstalk levels at output port 3 with optimized coupling length of 6 µm. Similar transmission curves for outputs 1 and 2 are expected for optimized coupling lengths of 4 µm and 4.5 µm, with simulated crosstalks <-30 dB and <-25 dB respectively.

We simultaneously launch a single 10-Gb/s data channel into all the 3 input ports of the mode multiplexer and measure a small power penalty (less than 1.9dB for BER of $10^{-9}$) on each output port of the mode demultiplexer. The experimental setup for performance evaluation is illustrated in Fig. 3a. In order to measure these power penalties the laser channel at 1563 nm is modulated with PRBS $2^{31}$-1 on-off-keyed (OOK) data by an amplitude modulator and then further phase-imprinted with a swept-frequency sinusoid in order to enable bit-error-rate (BER) measurements on channels which experience coherent crosstalk (further information provided in the Methods section). The data signal is then amplified, split evenly between the 3 input ports of the on-chip mode multiplexer, and simultaneously injected in quasi-TE polarization to the multiplexer ports. The varying fiber spans leading to the device ensure that the data is decorrelated between the ports (see methods section). The demultiplexed signals are recovered one at a time for inspection on a DCA and BER evaluation. Error free transmission (BER < $10^{-12}$) and open eye diagrams (Fig. 3b,c) are observed for all output ports. To account for fabrication imperfections, we improve the performance of port 3 at the expense of increased crosstalk and spectral filtering penalties on port 2 by wavelength detuning the $TE_1$ multiplexer ring. This enables device operation with an overall balanced power penalty (measured at a BER of $10^{-9}$) of 1.9 dB on ports 2 and 3 and 0.5 dB on port 1. In order to verify that intra-channel crosstalk is indeed the main mechanism of signal degradation[31], we also inspect the channel performance with only one input port injected at a time. We observe that transmission penalties result in 0.1 dB penalties on ports 1 and 2 and 0.8 dB on port 3 (with the higher penalty on port 3 resulting from the higher insertion loss through this port which leads to a larger OSNR degradation at the post-chip EDFA). Therefore we conclude that crosstalk is the main contributing factor to signal degradation in this device. The penalties are measured relative to a back-to-back (B2B) reference case which is defined and measured by replacing the chip with a tunable attenuator set to replicate the fiber-to-fiber loss of the lowest insertion-loss port.

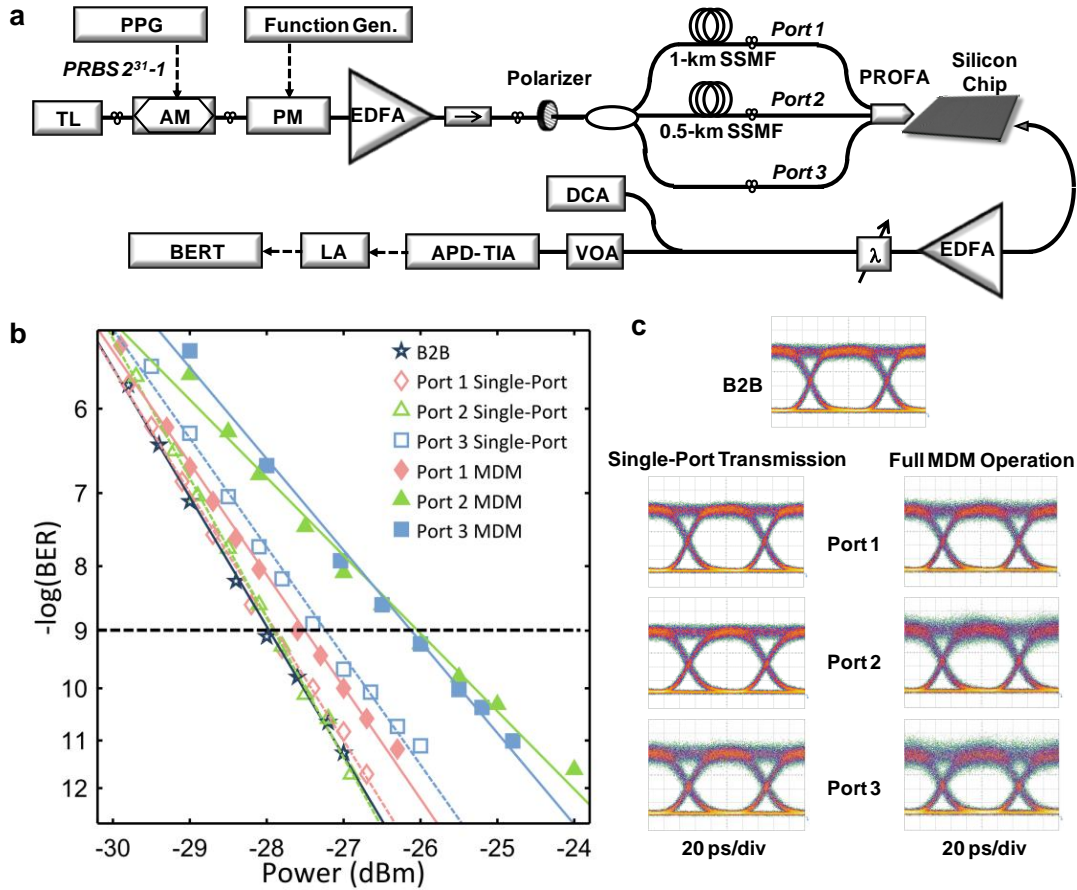

**Figure 3 | 3 Mode x 10 Gb/s Multiplexing Link. a.** Experimental setup for performance evaluation including Pulsed Pattern Generator (PPG), Tunable Laser (TL), Amplitude Modulator (AM), Phase Modulator (PM), Erbium-Doped Fiber Amplifier (EDFA), Isolator (→), Standard Single Mode Fiber (SSMF), Tunable Filter (λ), Digital-Communications Analyzer (DCA), Variable Optical Attenuator (VOA), Avalanche-Photodiode (APD-TIA), Limiting Amplifier (LA), and Bit-Error-Rate Tester (BERT). **b.** BER measurements for back-to-back (B2B) test case, single port transmission, and full MDM operation for all 3 ports. **c.** Corresponding eye-diagrams for the inspected signals.

We measure a low (< 1.4 dB) power penalty for joint MDM-WDM operation by launching three different 10-Gb/s wavelength channels spanning the full C-band into two input ports of the multiplexer (ports 1 and 2). A modified setup (depicted in Fig. 4a) is used to correctly decorrelate the wavelength channels (see methods section) and more polarizers are included in it to ensure that all the wavelength channels are launched on chip at the quasi-TE polarization with equal power. We set the wavelength channels to span the full C-band (limited by the EDFA gain band) and the microrings are tuned on-resonance to maximize power transmission at 1547 nm. The power penalties for both ports vary between 0.6 and 1.4 dB for all three wavelength channels (Fig. 4b) with performance variation attributed to slightly varying levels of crosstalk for the different wavelength channels. Error free transmission (BER < $10^{-12}$) and open eye diagrams (Fig. 4c) are observed for all the three channels at the two output ports. These results show that only a minimal penalty is added by extending the device operation to support WDM concurrently with the MDM.

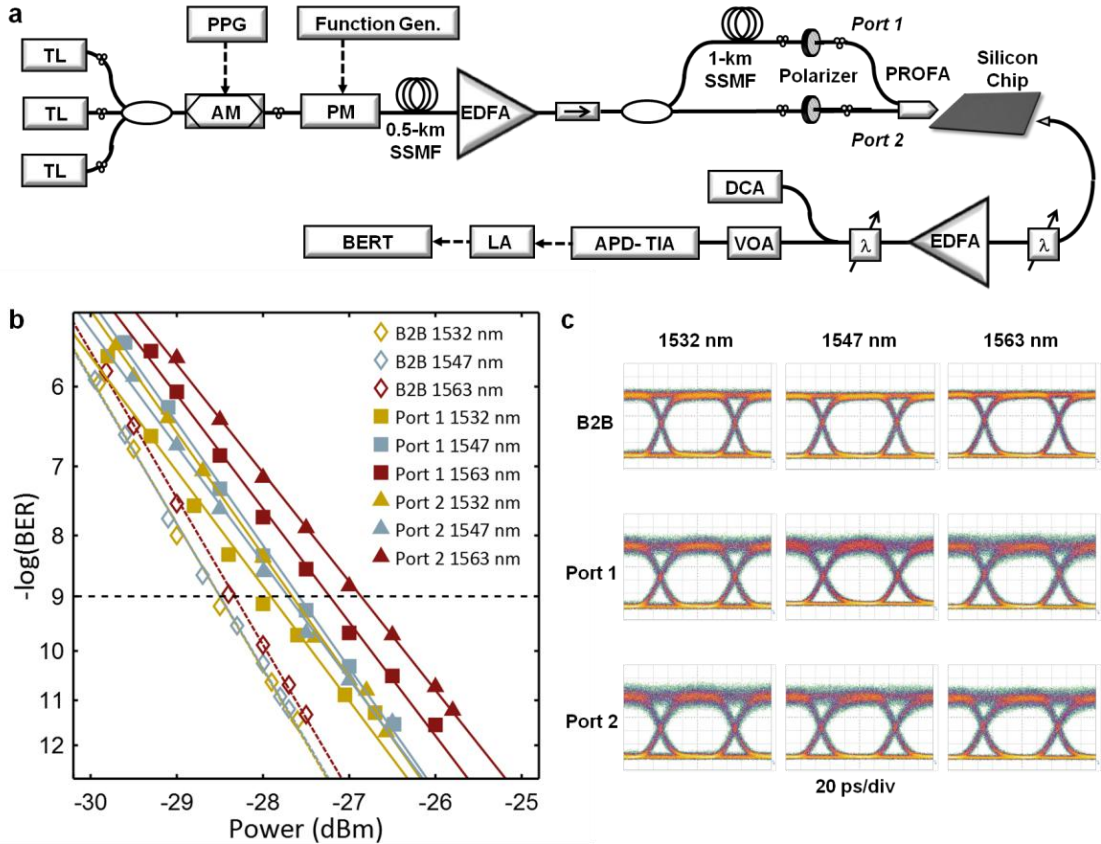

**Figure 4 | 2 mode x 3 wavelength x 10 Gb/s MDM-WDM link. a.** Experimental setup for demonstrating combined MDM and WDM operation. **b.** BER measurements for back-to-back (B2B) test cases and full MDM-WDM operation for both ports. **c.** Corresponding eye-diagrams for the inspected signals.

In conclusion, we have shown a platform that enables on-chip MDM-WDM optical interconnection for ultra-high bandwidth communications. Our simulation in Fig. 1a showed that when the multimode bus waveguide width tapers up to 2.37 µm, 5 spatial modes can be supported by this platform. In principle using a wider waveguide one could support an even larger number of modes. Each microring resonator in practice is able to support 87 WDM channels over the entire C-band (1530 – 1565 nm) by increasing the microring size such that the channel spacing is 50 GHz. Therefore, the on-chip MDM-WDM platform with the above-mentioned dimensions can potentially support an aggregate data rate up to 4.35 Tb/s with 5 spatial modes and 87 WDM channels.

**Methods**

**Device design and fabrication** – We fabricate the reconfigurable MDM-WDM silicon microring resonators on a 250 nm SOI wafer with 3 µm of buried oxide using standard CMOS fabrication processes. We patterned the waveguides using e-beam lithography with the dimensions as followed: the input/output ports and the microring are 450 nm wide; the $TE_0$, $TE_1$, and $TE_2$ (de)multiplexers have a width of 450 nm, 930 nm, and 1.41 µm wide; and each (de)multiplexer is linked by an adiabatic taper of 80 µm long. All the microrings have a radius of 10 µm and a coupling length of 5 µm. The separation gap between the microrings and all the input waveguides is 240 nm while the separation gap between the microrings and $TE_0$, $TE_1$, and $TE_2$ (de)multiplexers are 240 nm, 200 nm and 200 nm respectively. The silicon waveguides are then etched, followed by the e-beam resist being stripped and the etched structures are clad with 1 µm thick silicon oxide layer using plasma enhanced chemical vapor deposition to confine the optical mode. 300 nm of NiCr are next evaporated on the microrings resonators above the cladding to create the 1 µm wide heaters. Finally, 500 nm of gold (Au) are evaporated to define the electrical wires and contact pads using a lift-off process. The final fabricated device is illustrated in Fig. 2a. The footprint of this device is 0.11 $mm^2$ excluding the electrical wires.

**Optical modes imaging at the cross-section of the multimode waveguide** – We fabricate a test device that only has the mode multiplexer section ($TE_0$, $TE_1$, and $TE_2$) and is terminated with the 1.41-µm wide multimode bus waveguide. We couple a 1547 nm laser (on-resonance of each microring) into one of the input ports at one time. The output spatial modes of the multimode waveguide are then magnified with a 40x aspheric lens and imaged on an IR camera as shown in Fig. 2b.

**Phase dithering** – Intra-channel crosstalk[31] results in coherent interference of the laser with itself. In a test setup not employing any phase decoherence mechanisms, this results in a slow change of the output signal power as the phases leading to the device under test change as a result of thermal fluctuations in the fibers. If this remains untreated, these power fluctuations (on the temporal order of multiple seconds) prevent accurate BER measurements over short time spans. In order to enable finite-time BER measurements, two mechanisms are employed simultaneously to average out the slow phase fluctuations: 1. The arms leading to the multiplexer input ports are decorrelated by at least 0.5 km SSMF. This is close to the 1-km coherence length of the 200-kHz linewidth lasers we used in the experiments therefore ensuring some phase decoherence of the signals. 2. In order to guarantee full phase orthogonality regardless of the intrinsic laser linewidth, we incorporate phase modulation of a repeating linearly chirped signal consisting of a frequency sweep from 20-MHz to 10-MHz over a 5-ms period. With a 0.5-km path difference (roughly 2.5 μs relative delay), the phase difference between adjacent ports oscillates over $2\pi$ at 5-kHz, guaranteeing averaging of the phase difference in power measurements averaged over 100 ms.

**Multi-port edge coupling** – employing a method similar to one previously reported[32] we couple to three input ports simultaneously using a Pitch Reducing Optical Fiber Array (PROFA) mounted on a fully angle adjustable stage. The PROFA alignment was optimized to be within 2-dB of the optimal coupling values for all the ports simultaneously. Output coupling is done with a tapered lensed fiber aligned to one output port at a time.

**Intra-channel crosstalk penalties** – the back-to-back reference test case for power penalty measurements is defined by bypassing the chip and emulating insertion loss for the lowest loss port (port 1) with a variable optical attenuator. Intra-channel crosstalk penalties result in a power penalty predicted analytically[31] as $PP = -10 \log_{10}(1 - 2\sqrt{\varepsilon})$ where $\varepsilon$ is the ratio of the desired-signal's power to interfering signals' powers.

**Channel decorrelation** – in order to guarantee correct characterization of the device the input data channels which originate from a single PPG need to be decorrelated. The data channels coupled to the device's ports in the first experiment are decorrelated by 0.5-km and 1-km long fiber delays which ensure the patterns are relatively shifted between ports by at least 24 kb out of the pattern length of 2 Gb ($2^{31}$-1 PRBS). In the second experiment the wavelength channels are first decorrelated using the dispersion of a 0.5-km fiber to achieve at least 90-bit relative delay between adjacent wavelength channels. The inputs to the two ports used in this experiment are also decorrelated by a fixed 1-km fiber delay which guarantees decorrelation between the ports used.

## Acknowledgements

The authors gratefully acknowledge support from the U.S. Air Force (AFOSR) program FA9550-09-1-0704 on "Robust and Complex on-chip Nanophotonics" supervised by Dr. Gernot Pomrenke, support from DARPA for award #W911NF-11-1-0435 supervised by Dr. Jagdeep Shah. This material is based upon work supported by the National Science Foundation under Grant No. 1143893. This work was supported in part by the NSF through CIAN ERC under Grant EEC-0812072 and was performed in part at the Cornell NanoScale Facility, a member of the National Nanotechnology Infrastructure Network, which is supported by the NSF. The authors also acknowledge support from the NSF and Semiconductor Research Corporation under grant ECCS-0903406 SRC Task 2001. Lian-Wee Luo acknowledges a fellowship from Agency of Science, Technology and Research (A*STAR), Singapore.**Author Contribution**

L.W.L and N.O. contributed equally to this work. L.W.L. designed, and fabricated the device with the assistance of L.H.G. in the simulations. L.W.L, N.O. and C.C conducted the experiments together. L.W.L, N.O., K.B. and M.L. discussed the results and implications. L.W.L, N.O. and M.L. contributed to the writing of this paper.

**Competing financial interests**

The authors declare no competing financial interests.

**References**

1   Nagarajan, R. *et al.* Large-scale photonic integrated circuits. *Selected Topics in Quantum Electronics, IEEE Journal of* **11**, 50-65 (2005).

2   Jalali, B. & Fathpour, S. Silicon Photonics. *Lightwave Technology, Journal of* **24**, 4600-4615 (2006).

3   Richardson, D. J., Fini, J. M. & Nelson, L. E. Space-division multiplexing in optical fibres. *Nat Photon* **7**, 354-362 (2013).14

**Supplementary information**

Coupled-mode theory for two weakly coupled optical modes relates the complex amplitudes of the modes, $a_1$ and $a_2$ through a set of differential equations[1]:

$$\frac{da_1}{dz} = -j\beta_1 a_1 + \kappa_{12} a_2 \quad (1)$$

$$\frac{da_2}{dz} = -j\beta_2 a_2 + \kappa_{21} a_1. \quad (2)$$

The solutions to this equation set, assuming the waves $a_1(0)$ and $a_2(0)$ are launched at z = 0, are given by:

$$a_1(z) = \left[ a_1(0)\left( \cos\beta_0 z + j\frac{\beta_2 - \beta_1}{2\beta_0}\sin\beta_0 z \right) + \frac{\kappa_{12}}{\beta_0} a_2(0)\sin\beta_0 z \right] e^{-j[(\beta_1+\beta_2)/2]z} \quad (3)$$

$$a_2(z) = \left[ \frac{\kappa_{21}}{\beta_0} a_1(0)\sin\beta_0 z + a_2(0)\left( \cos\beta_0 z + j\frac{\beta_1 - \beta_2}{2\beta_0}\sin\beta_0 z \right) \right] e^{-j[(\beta_1+\beta_2)/2]z} \quad (4)$$

where

$$\beta_0 = \sqrt{\left(\frac{\beta_1 - \beta_2}{2}\right)^2 + \kappa_{12}\kappa_{21}}. \quad (5)$$

If the initial waves $a_1(0) = 1$ and $a_2(0) = 0$ are assumed, then the coupling from $a_1$ to $a_2$ is given by $\left|\frac{\kappa_{21}}{\beta_0}\sin\beta_0 z\right|$. Below we take a look at the example of the 1.41-μm wide multimode waveguide and investigate the coupling strength between the phase-matched $TE_2$ spatial mode and the $TE_0$ mode of 450-nm wide waveguide at λ = 1.55 µm (see Fig. 1a). The coupling gap between the two different waveguides is fixed at 200 nm. To achieve 100% power transfer from the 450-nm waveguide ($TE_0$) to the 1.41-µm waveguide ($TE_2$) using a directional coupler and vice versa, the coupling length ($L_{coupling}$) is required to be 42 µm (see Fig. 1c). To realize a compact device, we use microring resonators instead of a directional coupler. Weak coupling (achievable with a short coupling length) is sufficient to transfer nearly all the power from the input waveguide of the microring to the multimode waveguide (typically termed critical coupling)[2] (see Fig. 1b). We also calculate the coupling strength of the unwanted modes ($TE_1$, $TE_0$) which determines the crosstalk of the device. We observe that there is coupling to the undesired modes ($TE_1$, $TE_0$), however the coupling is weak due to the phase mismatch between these modes. The maximum coupling strength of $TE_1$ mode is 0.057. The optimum operating regime for low crosstalk is at $L_{coupling}$ ≈ 6 µm where the coupling to the undesired modes ($TE_1$, $TE_0$) is minimized.

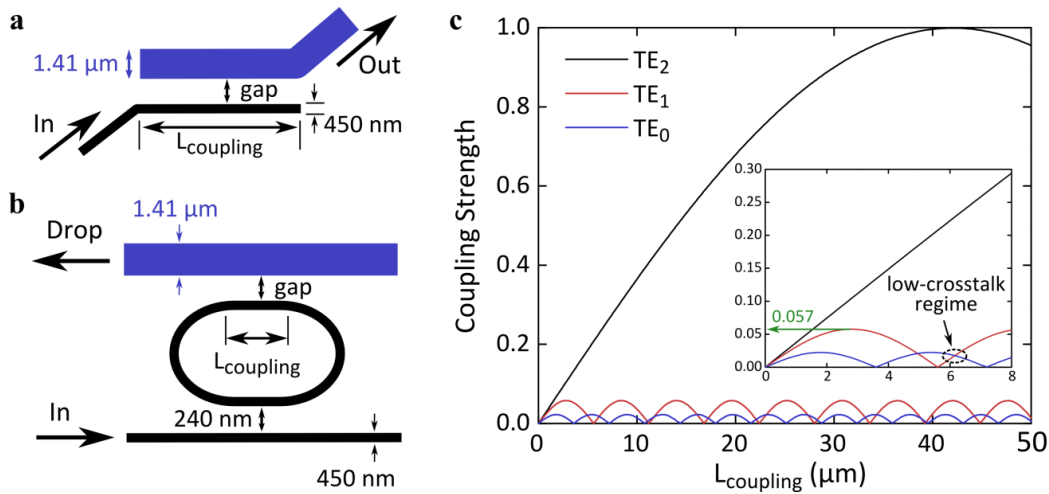

**Figure 1 | Coupling strength of different spatial modes ($TE_2$, $TE_1$, $TE_0$) of a 1.41-µm wide waveguide to the $TE_0$ mode of a 450-nm wide silicon waveguide with a coupling gap of 200 nm.**

**a.** Schematic of the coupling between a 1.41-μm wide waveguide and a 450-nm wide silicon waveguide. **b.** Schematic of an add-drop microring with asymmetric input and drop waveguides. **c.** The maximum coupling strength of $TE_1$ mode is 0.057. The optimum operating regime for low-crosstalk regime is at $L_{coupling} \approx 6$ μm.

The crosstalk can be further reduced by introducing weaker coupling at the microring-multimode waveguide coupling region (by having a larger coupling gap) to lower the maximum coupling of the undesired modes at the expense of longer coupling length to maintain the critical coupling condition. Figure 2 shows the coupling strength of the same microring-waveguide except that the coupling gap is changed to 280 nm. This weaker coupling at the coupling region lowers the maximum coupling strength of the $TE_2$ mode from the initial value of 0.057 to 0.037. This in turn results in a longer coupling length to maintain the critical coupling condition.

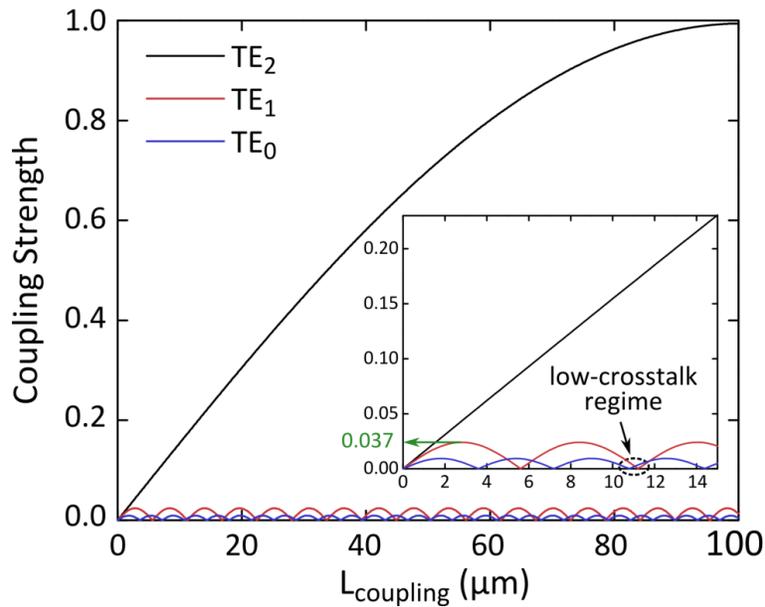

**Figure 2 | Coupling strength of different spatial modes ($TE_2$, $TE_1$, $TE_0$) of a 1.41-μm wide waveguide to the $TE_0$ mode of a 450-nm wide silicon waveguide with a coupling gap of 280 nm.**

The maximum coupling strength of $TE_1$ mode is 0.037. The optimum operating regime for low-crosstalk regime is at $L_{coupling} \approx 11$ μm.